# Making AI Philosophical Again: On Philip E. Agre's Legacy

Jethro Masís

*"So here I was in the middle of the AI world—not just hanging out there but totally dependent on the people if I expected to have a job once I graduated—and yet, day by day, AI started to seem insane. This is also what I do: I get myself trapped inside of things that seem insane."* — Philip E. Agre, RRE News and Recommendations (7/12/2000)

## Introduction

Philip Agre is a former professor of information studies at the University of California, Los Angeles, who was reported missing in October 2009. Concerned missing notes appeared widely on the internet (Carvin 2009; Pescovitz 2009; Young 2009; Rothman 2009) for he was presumed to have had some kind of mental breakdown and to be homeless somewhere in Los Angeles. In January 2010 it was reported that he was found alive and indeed "self-sufficient and in good health" (Carvin 2010). However, he never returned to academic life and explicitly asked his closest friends to be left alone. And this is also how the work of one of the most interesting theorists on the relationship between the computer and information sciences and the humanities faded away. He was an internet scholar and a sort of proto-inventor of web 2.0 (in the 1990s!) who grew increasingly worried about the consequences that communication technologies were having on people's privacy. Agre theorized this with his usual shrewdness and theoretical discernment (Agre 1994; Agre 1998). People who 'knew' him, or at least who worked close to him, describe Agre as a very reclusive person who never really spoke about his personal life. As reported by *The Chronicle of Higher Education* (Young 2009), Agre had not just gone missing but had abandoned his job and his apartment and he suffered from manic depression.

When Agre was still missing, Michael Travers—a computer scientist who met him in Graduate School at MIT—summarized in a blog post (Travers 2009) Agre's significance for computer studies and beyond. I think his words are worth quoting at length:

*Phil was one of the smartest people I knew in graduate school. More than smart, he had the intellectual courage to defy the dominant MIT sensibilities and become not just an engineer but also a committed critic of the ideology under the surface of technology, especially as it was applied*

 

*to artificial intelligence. He was a leader of the situated action insurgency in AI, a movement that questioned the foundations of the stale theories of mind that were implicit in the computational approach. Phil had the ability to take fields of learning that were largely foreign to the culture of MIT (continental philosophy, sociology, anthropology) and translate them into terms that made sense to a nerd like me. I feel I owe Phil a debt for expanding my intellectual horizons.*

*Phil was a seminal figure in the development of Internet culture. His Red Rock Eater email list[1] was an early predecessor to the many on-line pundits of today. Essentially he invented blogging, although his medium was a broadcast email list rather than the web, which didn't yet exist. He would regularly send out long newsletters containing a mix of essays, pointers to interesting things, and opinions on random things. He turned email into a broadcast medium, which struck me as weird and slightly undemocratic at the time, but he had the intellectual energy to fuel a one-man show, and in this and other matters Phil was just ahead of the times.*

This paper discusses some ideas envisioned by Agre, particularly the ones concerning a critical technical practice and the possibilities of making this practice (AI, for instance) philosophical again. Although I have Agre's work in high regard, I shall criticize his idea that finding the right technical implementation for everyday practice can be achieved under the rubric of programming Heidegger's *Zuhandenheit*. I am also appreciative of Agre's idea that there are certain metaphors pervading technical work which must be taken into account, but I will also argue that the Heideggerian so-called *Sichöffnende* and *Offene*, that is, the open character of human existence, is precisely not amenable to programming and that Agre could not get rid of the modern "technical construction of the human being as machine" (Heidegger Zoll, p. 178).

**Making AI Philosophical Again**

Philip Agre received his doctorate in Electrical Engineering and Computer Science at MIT, but he was always more interested (than many of his peers, that is) in exploring the philosophical assumptions pervading technological practices.[2] Thus, he deemed 'mistaken' to consider the Cartesian lineage of AI ideas as merely incidental and as having no purchase on the technical ideas that descend from it (1997, p. 23). Quite on the contrary, argues Agre, "computer systems are thus, among other things, also philosophical systems—specifically, mathematized philosophical systems—and much can be learned by treating them in the same way" (1997, p. 41). Agre claims "AI is philosophy underneath" (2005, p. 155); an assertion he clarifies in five points:

- AI ideas have their genealogical roots in philosophical ideas.
- AI research programs attempt to work out and develop the philosophical systems they inherit.
- AI research regularly encounters difficulties and impasses that derive from internal tensions in the underlying philosophical systems.
- These difficulties and impasses should be embraced as particularly informative clues about the nature and consequences of the philosophical tensions that generate them.
- Analysis of these clues must proceed outside the bounds of strictly technical research, but they can result in both new technical agendas and in revised understandings of technical research itself. (*idem*)

Influenced heavily by Dreyfus's pragmatization of Heidegger, Agre too understands *Sein und Zeit* as providing a phenomenology of ordinary routine activities, and believes Heidegger's *Analytik des Daseins* can provide useful guidance for the development of computational theories of interaction. Most importantly, it can also contribute to afford technical practice a historical conscience it overtly lacks, since "research like Heidegger's can have its most productive influence upon AI when AI itself recovers a sense of its own historical development" (Agre 1996, p. 25). This last critical and historical trait permits Agre, as a philosoher of computing, to denounce that modern computational practices can be viewed as the resolute incarnation of a disembodied conception of philosophy having Augustine, Descartes, and Turing as pivotal figures, with the opposition of body and soul at the core of their thinking:

*Each man's cultural milieu provided fresh meaning for this opposition: Augustine struggled to maintain his ideals of Christian asceticism,*

*Descartes described the soldier's soul overcoming his body's fear as the Thirty Years' War raged, and Turing idealized disembodied thought as he suffered homophobic oppression in modern England (1997, p. 103).*

This means, for Agre, that there is a historical tradition and discourse sustaining the practices of contemporary computational approaches, so by no means can they be said to sustain themselves exclusively on technical terms. The latter view is not only naïve but also dishonest. But unfortunately, Agre sees that computer science is utterly oblivious to "its intellectual contingency and recast itself as pure technique" (*idem*). This is the reason why Agre castigates this forgetfulness of the assumptions running deep in AI, which more often than not are compensated for, put aside, and substituted by the formalist attempt to cleanse computational programs of the 'inexactness' of natural language, and to strip AI altogether of its historical and cultural underpinnings. It is by virtue of not paying attention to how their scientific practices are constituted that formalists attempt to liberate computational work precisely from the unruliness and imprecision of vernacular language, which appears foreign and annoying to their technical field. Moreover, "they believed that, by defining their vocabulary in rigorous mathematical terms, they could leave behind the network of assumptions and associations that might have attached to their words through the sedimentation of intellectual history" (Agre 2002, p. 131). This is why Agre believes such an attempt should not be countenanced any longer but rather it should be confronted by means of a 'critical technical practice:' the kind of critical stance that would guide itself by a continually unfolding awareness of its own workings as a historically specific practice (1997, p. 22). As such, "it would accept that this reflexive inquiry places all of its concepts and methods at risk. And it would regard this risk positively, not as a threat to rationality but as a promise of a better way of doing things" (Agre 1997, p. 23).

This critical technical practice proposed by Agre has clear overtones oscillating amid an immanent critique, on the one hand, and an 'epistemological electroshock therapy' toward situating scientific knowledge (Haraway 1988), on the other. Indeed, Agre's view might appear problematic if it is ill-construed as confusing or ambiguous, since it can be seen both as a critique *from within*—accepting the basic methodology and truth-claims of computer science, peppered with internal disputes against the more obviously invalid and politically loaded claims—and as a critique *from without*—recognizing the ultimate cultural contingency of all claims to scientific truth (Sengers 1995, p. 151). Nevertheless, it is crucial for Agre to present his work as neither an internalist account *of* AI, nor as a philosophical study *about* AI, but as "actually a work *of* AI: an intervention within the field that contests many of its basic ideas while remaining fundamentally sympathetic to computational modeling as a way of knowing" (1997, p. xiv). Agre finds it more daring to intervene in the field and show practically how critical technical views might help develop better artificial systems. Both the critical intervention on the field and the fundamental sympathetic posture deserve, furthermore, a separate explanation.

With regard to the critical intervention, Agre notes that there is a certain mindset when it comes to what 'computer people' believe—and this is, of course, Agre's niche—regarding the aims and scope of their own work. This belief is not at all arbitrary or merely capricious, but rather it must be viewed in conjunction with the very nature of computation and computational research in general. According to Agre, computational research can be defined as an inquiry into physical realization as such. Moreover, "what truly founds computational work is the practitioner's evolving sense of what can be built and what cannot" (1997, p. 11). The motto of computational practitioners is simple: *if you cannot build it, you do not understand it. It must be built and we must accordingly understand the constituting mechanisms underlying its workings.* This is why, on Agre's account, computer scientists "mistrust anything unless they can nail down all four corners of it; they would, by and large, rather get it precise and wrong than vague and right" (1997, p. 13). There is also a 'work ethic' attached to this computationalist mindset: *it has to work*. However, Agre deems it too narrow to entertain just this sense of 'work.' Such conception of what counts as success is also ahistorical in that it can simply be defined as working because the program conforms to a pre-given formal-mathematical specification. But an AI system can also be said to work in a wholly different sense: when its operational workings can be narrated in intentional terms by means of words whose meaning goes beyond the mathematical structures (which is, of course, a pervasive practice in cognitive scientific

Jethro Masís
Making AI Philosophical Again: On Philip E. Agre's Legacy

explanations of mechanism). For example, when a robot is said to 'understand' a series of tasks, or when it is proclaimed that AI systems will give us deeper insights about human thinking processes. This is indeed a much broader sense of 'work,' one that is not just mathematical in nature, but rather a clearly discursive construction. And it certainly bears reminding that such discursive construction is part of the most basic explanatory desires of cognitive science. So in the true sense of the words 'build' and 'work,' AI is not only there to build things that *merely* work. Let us quote at length:

*The point, in any case, is that the practical reality with which AI people struggle in their work is not just 'the world,' considered as something objective and external to the research. It is much more complicated than this, a hybrid of physical reality and discursive construction. The trajectory of AI research can be shaped by the limitations of the physical world—the speed of light, the three dimensions of space, cosmic rays that disrupt memory chips—and it can also be shaped by the limitations of the discursive world—the available stock of vocabulary, metaphors, and narrative conventions. (Agre 1997, p. 15)*

This also gives hints as to how exogenous discourses, like philosophy, are supposed to be incorporated into technological practices. Agre is of the opinion that the point is not to invoke Heideggerian philosophy, for example, as an exogenous authority thus supplanting technical methods: "the point, instead, is to expand technical practice in such a way that the relevance of philosophical critique becomes evident *as a technical matter*. The technical and critical modes of research should come together in this newly expanded form of critical technical consciousness" (1997, p. xiii). The critical technical practice Agre envisions is one "within which such reflection on language and history, ideas and institutions, is part and parcel of technical work itself" (2002, p. 131). More exactly, Agre confesses that his intention is "to do science, or at least something about human nature, and not to solve industrial problems" (1997, p. 17). And he adds: "but I would also like to benefit from the powerful modes of reasoning that go into an engineering design rationale" (*idem*). In such a way, Agre pretends to salvage the most encompassing claims of AI research—that it can teach us something about the world and about ourselves—by means of incorporating a self-correcting, history-laden approach combining both technical precision and philosophical rigor. By expanding the comprehension of the ways in which a system can work, "AI can perhaps become a means of listening to reality and learning from it" (Agre 2002, p. 141). But it is because of its not listening to reality that, for instance, Dreyfus (1992) launched his attacks against AI as an intellectual enterprise.

On this account, Agre contends that merely "lashing a bit of metaphor to a bit of mathematics and embodying them both in computational machinery" (1997, p. 30)—which is usually what computer scientists come up with—will not do the job of contributing to the understanding of humans and their world. So framed, the approach appears to Agre as too narrow, naïve, and a clear way of not listening to reality. So he has a more ambitious project: the very metaphors being lashed to a bit of mathematics that end up in machinery implementation must be investigated. Both physical reality and discursive construction must be taken into account. Although technical languages encode a cultural project of their own (the systematic redescription of human and natural phenomena within the limited repertoire of technical schemata that facilitate rational control)—a fact which, incidentally, tends to be as such elided—"it is precisely this phenomenon that makes it especially important to investigate the role of metaphors in technical practice" (Agre 1997, p. 34). At this juncture, Agre sounds strikingly similar to Blumenberg, whose metaphorological project "seeks to burrow down to the substructure of thought, the underground, the nutrient solution of systematic crystallizations; but it also aims to show with what 'courage' the mind preempts itself in its images, and how its history is projected in the courage of its conjectures" (2010, p. 5). For Agre too, metaphors play a role in organizing scientific inquiry or, to say it with Blumenbergian tones, metaphors are by no means 'leftover elements' (*Restbestände*) but indeed 'foundational elements' (*Grundbestände*) of scientific discourse.[3] Clinging to Kuhnian terminology (see Kuhn 1996), this can also be couched in terms of the tension between normal science—with its aseptic attitude toward reducing instability of meaning and inconsistency via a cleansing of elements of inexact, ambiguous nature—and revolutionary science which makes metaphoric leaps that create new meanings and applications that might constitute genuine theoretical progress (Arbib & Hesse 1987, p. 157). By showing how technical practice is not only the result of technical work but also of discursive construction and unexplained metaphors, Agre's critical technical practice might meet the criteria for being considered a truly revolutionary approach in Kuhnian terms. It remains to be seen, however, whether this is indeed the case.

The sympathetic attitude towards computational modeling that Agre espouses takes as its point of departure the analysis of agent/environment interactions which accordingly should be extended to include the conventions and invariants maintained by agents throughout their activity. This notion of environment is referred to, with clear Husserlian overtones (see Husserl 1970), as *lifeworld,* and can be incorporated into computational modeling via "a set of formal tools for describing structures of lifeworlds and the ways in which they computationally simplify activity" (Agre & Horswill 1997, p. 111). From this follows that Agre's theoretical emphasis lies on the concept of *embedding*. This means that agents are not only to be conceived of as embodied but more crucially as embedded in an environment. The distinction between embodiment and embedding can be explained as follows:

*'Embodiment' pertains to an agent's life as a body: the finiteness of its resources, its limited perspective on the world, the indexicality of its perceptions, its physical locality, its motility, and so on. 'Embedding' pertains to the agent's structural relationship to its world: its habitual paths, its customary practices and how they fit in with the shapes and workings of things, its connections to other agents, its position in a set of roles or a hierarchy, and so forth. The concept of embedding, then, extends from more concrete kinds of locatedness in the world (places, things, actions) to more abstract kinds of location (within social systems, ecosystems, cultures, and so on). Embodiment and embedding are obviously interrelated, and they each have powerful consequences both for agents' direct dealings with other agents and for their solidarity activities in the physical world. (Agre & Horswill 1997, 111-112)*

The importance for cognitive science of having a well-developed concept of the environment is not to be underestimated, since it seems that only prior to a basic understanding of an agent's environment can a given pattern of adaptive behavior be figured out. Taking a stride towards defining the environment with at least a modicum of rigor amounts to developing "a *positive theory* of the environment, that is, some kind of principled characterization of those structures or dynamics or other attributes of the environment *in virtue of which* adaptive behavior is adaptive" (Agre & Horswill 1997, p. 113). Accordingly, Agre and Horswill lament that AI has downplayed the distinction between agent and environment by fatally reducing the latter to a discrete series of choices in the course of solving a problem, but "this is clearly a good way of modeling tasks such as logical theorem-proving and chess, in which the objects being manipulated are purely formal" (*idem*). AI can go on well without a well-developed concept of the environment but only at the price of focusing on mere toy-problems, microworlds, and toy-tasks within such artificial environments. It should then not come as a surprise that the situation changes dramatically for tasks involving physical activities, where "the world shows up, so to speak, phenomenologically: in terms of the differences that make a difference for*this* agent, given its particular representations, actions, and goals." (*idem*). The environmental indexicality that is brought forward here is often objected to by cognitivists as though agents performed tasks without any computation whatsoever, or as though agents inhabiting a lifeworld lived in an adamant reactive mode. But the point is rather that "the nontrivial cognition that people do perform takes place against a very considerable background of familiar and generally reliable dynamic structure" (Agre & Horswill 1997, p. 118). Now, precisely indexicality has been difficult to accommodate within AI research. With this in view, Agre has criticized the usual assumptions of the received view in technical practice as follows:

- That perception is a kind of reverse optics building a mental model of the world by working backward from sense-impressions, inferring what in the world might have produced them.
- That action is conducted through the execution of mental constructs called plans, understood as computer programs.
- And finally, that knowledge consists in a model of the world, formalized in terms of the Platonic theory analysis of meaning in the tradition of Frege and Tarski. (2002, p. 132)

The dissociation of mind and body (the founding metaphor of cognitive science and modern philosophy) is here at work, precisely when traditional AI thinks of the mind roughly as a plan generator and the body as the executor of the plan. Moreover, AI is so framed in terms of a series of dissociations: mind versus world, mental activity versus perception, plans versus behavior, the mind versus the body, and abstract ideas versus concrete things (Agre 2002, p. 132). According to Agre, these dissociations are contingent and can be considered 'inscription errors' (Smith 1996): "inscribing one's discourse into an artifact and then turning around and 'discovering' it there" (Agre 2002, p. 130). And this is not to be admired. As Nietzsche contented in *Über Wahrheit und Lüge im außermoralischen Sinne* (1873), when someone hides something behind a bush and looks for it again in the same place and finds it there as well, there is not much to praise in such seeking and finding.

That AI research has been framed along these contingent oppositions makes it clear that it is part of the history of Western thought. As such,

*it has inherited certain discourses from that history about matters such as mind and world, and it has inscribed those discourses in computing machinery. The whole point of this kind of technical model-building is conceptual clarification and empirical evaluation, and yet AI has failed either to clarify or to evaluate the concepts it has inherited. Quite the contrary, by attempting to transcend the historicity of its inherited language, it has blunted its own awareness of the internal tensions that this language contains. The tensions have gone underground, emerging through substantive assumptions, linguistic ambiguities, theoretical equivocations, technical impasses, and ontological confusions. (Agre 2002, p. 141)*

Nevertheless, it is interesting to note that—for all his philosophical acumen—Agre himself has not been able to liberate himself from the persistence of a representational theory of cognition, even when his is certainly more concrete, more historically conscious, and more enactive than the one customarily held in the traditional view. As a result, the latter critical concepts grouped together conform with the motivation for developing a concept of indexical-functional or deictic representation (Agre & Chapman 1987; Agre 1997), the main idea being that agents represent objects in generic ways through relationships to them (Agre & Horswill 1997, p. 118). On Agre's view, what must be done is refine the concept of representation (and not just cast it aside) and show what kind of representational activity is at work in interaction. Thus, the point is to criticize the underlying view of knowledge presupposed by the traditional theory of representation (that knowledge is picture, copy, reflection, linguistic translation, or physical simulacrum of the world), while suggesting that "the primordial forms of representation are best understood as facets of particular time-extended patterns of interaction with the physical and social world" (Agre 1997, p. 222). Therefore, "the notion of representation must undergo painful surgery to be of continued use" (Agre 1997, p. 250). Given that this redefinition of representation by Agre has its own quirks, it must now be carefully explained.

The traditional theory of representation, which has been put into work and is thoroughly presupposed in traditional AI research, is based on the notion of world model. Such notion refers to some structure that is thought to be within the mind or machine that represents the outside world by standing in a systematic correspondence with it (Agre 1997, p. 223). As such, the assumption that there is a world model being represented by the mind is the epitome of mentalism (Agre 1997, p. 225). Mentalism was previously defined by Agre as the generative metaphor pervasive in cognitive science according to which every human being has an abstract inner space called a 'mind' which clusters around a dichotomy between outside and inside organizing a special understanding of human existence (Agre 1997, p. 49). Marres (1989), a defender of mentalism, defines it as the view that the mind directs the body. Thus, on Agre's terms, giving preeminence to indexicality amounts to inverting this picture, since conceding that human beings are not minds that control bodies implies that interaction cannot be defined "in terms of the relationships among a mind, a body, and an outside world" (1997, p. 234), which is unfortunately so typical in cognitive scientific explanations. And here the key term is indeed interaction, understood not as the relation between the subjective and the objective, but rather as emerging from the actual practices people employ to achieve reference *in situ*. Indexicality "begins to emerge not merely as a passive phenomenon of context dependence but as an active phenomenon of context constitution" (Agre 1997, p. 233).

David Chalmers could only table the question *what is it like to be a thermostat?* (1996, p.

293)—thus blurring in one fell swoop the difference between living organisms and mechanism—by means of importing some heavy philosophical baggage, namely the assumption that the thermostat controls the temperature of systems in general (not of *this* specific system, say the internal combustion engine of a specific car), or that a thermometer measures the temperature "in room 11" (instead of *here*), or that one eats with "fork number 847280380" in some cosmic registry of forks (instead of *precisely this* fork I am holding with my left hand). Quite on the contrary, when indexicality is introduced as a constituting factor of interaction, it turns out that "human activities must be described in intentional terms, as being *about* things and *toward* things, and not as meaningless displacements of matter. Physical and intentional description are not incomparable, but they *are* incommensurable" (Agre 1997, p. 245). From this follows that the typical ascription of intentional states to nonembedded systems is absurd, precisely because embedding, and the interaction deriving thereof, is the condition of possibility of intentional comportment. The ubiquitous character of experience suggested in Chalmers's classic book on the philosophy of consciousness is also an inscription error, for it arises from obviating the need for a proper theory of intentionality or, to be more exact, such view derives from the naturalization of intentionality. When actual, concrete intentional activities are taken off the picture, representation is no longer connected with a lifeworld. Thus, the illusion can be then entertained that a semantic theory merely entails the categorization in some objective way of the ontology of a concrete situation, before the event of activity has taken place, or ignoring *tout court* the eventual character of activity (Agre 1997, p. 232).

Incidentally, the aforementioned illusion is Agre's critique of the semantic theory espoused by Barwise and Perry (1983), which, on Agre's criticism, comports a metaphysical realism that obscures indexicality. According to Agre, "when a speaker uses an indexical term such as 'I,' 'you,' 'here,' 'there,' 'now,' or 'then' to pick out a specific referent, this picking out is determined by relations between situations; it is not an *act* on the speaker" (1997, p. 233). These interactions and how they shape situations must be clarified, since it can be said that "interaction is central, both to human life and to the life of any agent of any great complexity" (Agre 1997, p. 234). Embedded activities must be investigated in how they are structured, as well as the sort of representing which is most incumbent on them.

For Agre, the latter requires a proper theory of intentionality couched within the Heideggerian distinction between *Zuhandenheit* and *Vorhandenheit* (SZ § 15). Traditional AI research can be accused of having only paid attention to present-at-hand phenomena, thus attempting to model computationally what precisely appears salient objectively in perception. In contrast, Agre finds that this phenomenological distinction is neither psychological nor mechanistic but a description of the structure of everyday experience that can be suitable for a new way of computational modeling of that experience. Preston (1993) had already explored this Heideggerian distinction in relation to another one: that of nonrepresentational and representational intentionality. One could, *à la* Dreyfus (2002a; 2002b), identify respectively *Vorhandenheit* with representational intentionality and *Zuhandenheit* with a sort of nonrepresentational intentionality and so proclaim beforehand the failure of artificial systems propounding the accomplishment of high-level intelligence. For Agre, however, this is too radical and, above all, too pessimistic. What is needed is a clarification of what kinds of representation exist and the role they play in real activities (Agre 1997, p. 237). Herein resides the importance of delving into experience and providing AI with a set of tools to enrich its vocabulary and metaphors. This is needed because "the philosophy that informs AI research has a distinctly impoverished phenomenological vocabulary, going no further than to distinguish between conscious and unconscious mental states" (Agre 1997, p. 239). Agre is onto something more important here, which is nothing less than making AI philosophical again: "technology at present is covert philosophy; the point is to make it openly philosophical" (1997, p. 240).

The traditional idea of representation understood it as a model in an agent's mind that corresponds to the outside world through a systematic mapping. Agre opines that AI research has been concerned only with a partly articulated view of representation. No wonder, then, the meaning of representations for an agent can be determined almost as *en-soi*—to use Sartre's terminology in *L'être et le néant* (see Sartre 1984)—without any reference being provided as to the agent's location, attitudes, interests, and idiosyncratic perspective (as *être-pour-soi*). This is also the reason explaining why "indexicality has been almost entirely absent from AI research" (Agre 1997, p. 241). Moreover, "the model-theoretic understanding of representational semantics has made it unclear how we might understand the concrete relationships between a representation-owning agent and the environment in which it conducts its activities" (*idem*). On Agre's view, the reason why AI research has lagged behind a clear-cut understanding of representation and indexicality has not been its nondistinctiveness between mechanism and human phenomena. Notwithstanding Agre's crucial imports from the alien province of phenomenology, he would nevertheless defer to Chalmers's highly controversial idea that experience is ubiquitous, albeit with a caveat: the problem is not to ask whether there is something it is like for a thermostat to be what it is, for Agre has it that any device that engages in any sort of interaction with its environment can be said to exhibit some kind of indexicality (1997, p. 241). Chalmers's problem is simply not to have considered exactly which kind of intentionality might be ascribed to artifacts like thermostats. Artifacts do have some sort of ambience embedding. In example, "a thermometer's reading does not indicate abstractly 'the temperature,' since it is the temperature *somewhere*, nor does it indicate concretely 'the temperature in room 11,' since if we moved it to room 23 it would soon indicate the temperature in room 23 instead. Instead, we need to understand the thermometer as indicating 'the temperature *here*'—regardless of whether the thermometer's designers thought in those terms" (*idem*). As Agre's contention goes, the point is to ascribe indexicality to artifacts. In fact, "AI research needs an account of intentionality that affords clear thinking about the ways in which artifacts can be involved in concrete activities in the world" (1997, p. 242).

Such account of intentionality was coined by Agre under the rubric of deictic representation as opposed to objective representation. First, two sorts of ontology are to be distinguished. According to an objective ontology, individuals can be defined without reference to activity or intentional states. A deictic ontology, by contrast, can be defined only in indexical and functional terms and in relation to an agent's location, social position, current goals and interests, and autochthonous perspective (Agre 1997, p. 243). Entities entering the space of whatever interaction with the agent, can only be understood correctly in terms of the roles they play in the agent's activities. In accordance with the aforementioned deictic notation introduced by Agre, "some examples of deictic entities are *the-door-I-am-opening*, *the-stop-light-I-am-approaching*, *the-envelop-I-am-opening*, and *the-page-I-am-turning*. Each of these entities is indexical because it plays a specific role in some activity I am engaged in; they are not objective, because they refer to different doors, stop lights, envelopes, and pages on different occasions" (*idem*). Their nonobjective character, however, does not imply that, by contrast, indexical entities are to be considered as subjective and, for that matter, as phantasms or internal and intimate qualia. The idea behind this is precisely that a deictic ontology should not be confused with subjective, arbitrary musings of an encapsulated subject. In the first place, this is the ontology that can be most properly ascribed to routine activities. Therefore, it would be preposterous to suggest that they are private or ineffable. Routines and activities are realized 'out there' in the world and, for that very reason, do not pertain to an internal mental game: they are, indeed, public. Accordingly, in routine activities the objective character of entities with which one copes, is not salient or important. Neither is their 'subjective feel,' nor the way they appear to me as individual. That their character is deictic means that what is most important is the role they play in the whole of activity. Therefore, hyphenated noun phrases like *the-car-I-am-passing* or *the-coffee-mug-I-am-drinking-with* are not mental symbols in the cognitivist sense. They designate "not a particular object in the world, but rather a role that an object might play in a certain time-extended pattern of interaction between an agent and its environment" (Agre 1997, p. 251).

Agre's alternative way of conceiving of activity and the express purpose of modeling it computationally is very attractive. As a matter of engineering, the leading principle is that of machinery parsimony: "choosing the simplest machinery that is consistent with known dynamics" (Agre 1997, p. 246). This view explicitly contrasts with the emphasis on expressive and explicit representation typical of traditional AI, with all the inherent difficulties of programming beforehand, as scripts, all the situations an artificial agent might encounter when coping with the world. By clear contrast with traditional AI, "the principle of machinery parsimony suggests endowing agents with the minimum of knowledge required to account for the dynamics of its activity" (Agre 1997, p. 249). In such a way, Agre's approach also resonates with Brooksian tones (see

Issue 4.1 / 2014: 65

Jethro Masís  Making AI Philosophical Again: On Philip E. Agre's Legacy

continentcontinent.cc/index.php/continent/article/view/177

Brooks 1999) of removing 'intelligence' and even 'reason' from the picture in order to render an account of interactive representation. Moreover, Agre sees deictic representation as changing the traditional view altogether since it presents us with the possibility, not of expressing explicitly and in every detail objective states of affairs, but of *participating* in them: "conventional AI ideas about representation presuppose that the purpose of representation is to express something, but this is not what a deictic representation does. Instead, a deictic representation underwrites a mode of relationship with things and only makes sense in connection with activity involving those things" (1997, p. 253). However, the objection may be raised that such a deictic approach violates the grand spirit of AI which seeks greater explicitness of representation and broader generality, since Agre's formula for design might simply contribute to model only special-purpose—and thusly limited—devices. But Agre responds that "the conventional conception of general-purpose functionality is misguided: the kind of generality projected by current AI practice (representation as expression, thought as search, planning as simulation) simply cannot be realized" (1997, pp. 249-250).

This is, of course, not just a series of theoretical postulates urged by Agre, since he distinguishes amongst levels of analysis (1997, pp. 27-28). The *reflexive* level, which has been already exhibited in the previous pages of this exposition, provides ways for analyzing the discourses and practices of technical work. Given that technical language is unavoidably metaphorical, the reflexive level permits one to let those metaphors come to the surface and thus can they be taken into account when technical work encounters trouble in implementation. On the *substantive* level, the analysis is carried out with reference to a particular technical discipline, in this case AI. But Agre is primarily interested in proceeding, on top of the reflexive and substantive levels, on a *technical* level, in order to explore "particular technical models employing a reflexive awareness of one's substantive commitments to attend to particular reality as it becomes manifest in the evolving technical work" (1997, p. 28). On Agre's view, traditional AI practitioners have not conscientiously attended to this partitioning of levels of analysis. Particularly, the reflexive level that prescribes an awareness of the role of metaphors in technical work has been disdained, as though AI researchers could simply bootstrap their way to technical success without being aware of the underlying metaphors pervading their work. For Agre, this is particularly problematic because "as long as an underlying metaphor system goes unrecognized, all manifestations of trouble in technical work will be interpreted as technical difficulties and not as symptoms of a deeper, substantive problem" (1997, p. 260).

As an exemplary case of technical work based on the aforementioned levels of analysis, Agre presents Pengi, a program designed by Chapman and Agre (1987) in the late 1980s under the rubric of being an implementation of a theory of activity. Pengi is a penguin portrayed in the commercial computer game Pengo, who finds itself in a maze made up of ice blocks that is surrounded by an electric fence. The maze is also inhabited by deadly bees that are to be avoided at all costs by Pengi and the task of the player is to maintain Pengi alive and defend it from such perils coming along the way. As defense, the bees can be killed by crushing them with a moving ice block or by kicking the fence while they are touching it. This momentarily stuns the bees and they can be crushed by simply walking over them. Agre argues that Pengo is an improvement on the blocks world, although it obviously fails to capture numerous elements of human activity. What is important is the combination of goal-directedness and improvisation involved in the game, from which Agre hopes to learn some computational lessons. First of all, Agre and Chapman did not attempt to implement in advance everything they knew about the game, thus contradicting the mapping out beforehand which is typical in traditional AI systems. The point is to see Pengi as relating to the objects that appear in its world, not in terms of their resemblance to mental models which were beforehand programmed, but solely in terms of the roles they play in the ongoing activity. As such, what Agre and Chapman attempted to program was actually deictic representations:

- *the-ice-cube-which-the-penguin-I-am-controlling-is-kicking*,
- *the-bee-I-am-attacking*,
- *the-bee-on-the-other-side-of-this-ice-cube-next-to-me*, etc.

At any rate, Agre does not argue that this simple system can be regarded as intelligent: "Pengi does not understand what it is doing. No computer has ever understood to any significant degree what it was doing" (1997, p. 301). But the bottom line is straightforward enough to explain: the game constituting Pengi's world as agent is not made up of present-at-hand entities and processes, but more importantly of possibilities for action that require appropriate responses from the agent. This shows Agre's understanding of ready-to-hand entities as no entities at all, but as possibilities for action and subsequent responses to the demands of the situation at hand. Given that these possibilities for action are not objects at all and that usually this sort of open stance for responding skillfully to environmental challenges does not appear in propositional referring, it is understandable that they have been rather elusive for programmers. After all, how can one program possibilities for action, since the focus is not on this particular object or the other but rather on the movement constituting the towards-which for-the-sake-of-which? The wellspring of this movement is all the more elided because, as Heidegger has it, precisely what is closest to us ontically is ontologically (and for that very reason) that which is farthest (SZ § 5, p. 15). This has been Agre's task, namely: to attempt to reveal the ontological dimension by means of technological implementation that does not obfuscate it but that rather embraces it. By programming deictic representations instead of just objective ones, Agre argues, computational programs can learn this fundamental lesson: what was lacking in traditional AI systems was precisely a model to envision a specific relationship between machinery and dynamics based on the concept of interaction. This lesson, so the argument goes, can gradually dispel the need for mentalist approaches.

## Conclusion

It should be noted that Agre was clearly influenced by Dreyfus's early critique of artificial reason (see Dreyfus 1992) but his path was individually constructed and his insights were also supported by different motivations. What is perhaps more interesting is his willingness to go on and program something, and this after researching with seriousness the history of the philosophy of mind and drawing even upon continental sources. Dreyfus credits him with expounding the philosophical debate and even with understanding *Zuhandenheit* better than he himself did, since for Agre ready-to-hand is not a *what* but a *for-what* (2007, p. 252). Dreyfus has it that Agre was able to show how Heidegger wants to get at something more basic than simply a class of objects: equipment (*Zeug*). The entire point of the equipmental character of things in the world is not that they are entities with a function feature—this was Dreyfus's pre-Agrean interpretation of *Zeug* and *Zeugzusammenhang*—but rather that they open up possibilities for action, solicitations to act, and motivations for coping; an idea that Dreyfus takes admittedly from Agre's endeavors towards modeling *Zuhandenheit* on the basis of deictic intentionality. Nevertheless, Dreyfus is of the opinion that in attempting to program ready-to-hand, Agre succumbs to an abstract objectification of human practice, because affordances—inasmuch as they are not objects but the in-between interaction in which no subject nor object is involved—are not amenable to programming. That they are not is not something Agre seems to fully understand, and this is why he thinks that somehow deictic representations must be involved in human understanding. According to Dreyfus, "Agre's Heideggerian AI did not try to program this experiential aspect of being drawn in by an affordance. Rather, with his deictic representations, Agre *objectified* both the functions and their situational relevance for the agent. In *Pengi*, when a virtual ice cube defined by its function is close to the virtual player, a rule dictates the response (e.g., kick it). No skill is involved and no learning takes place" (2007, p. 253). It must be admitted that a virtual world is not even slightly comparable with the complex dynamics of the real world. In a virtual world, the dynamics of relevance are determined beforehand, so a program like *Pengi* simply cannot account for the way human beings cope with new relevancies. Dreyfus concludes that Agre "finesses rather than solves the frame problem. Thus, sadly, his Heideggerian AI turned out to be a dead end. Happily, however, Agre never claimed he was making progress towards building a human being" (2007, p. 253).

Agre's contribution consists in his attempt to program *Zuhandenheit* instead of *Vorhandenheit*. That this can be made is, however, highly controversial. Certainly, what is deeply contentious is not that phenomenological insights can be brought to bear on cognitive science for a critical technical practice like the one Agre requires, but rather the assumption that the experiential dimension which phenomenology has revealed can be programmable. According to Heidegger, "*the essence of Dasein lies in its existence*" (SZ § 9, p. 42), which does not imply any "'properties' present-at-hand of some entity which 'looks' so and so and is itself present-at-hand" (*idem*). To exist as Dasein, then, implies that one's own existence has to be partly constructed, for existence is not already given and therefore is no program that can be run by any kind of hardware (Capurro 2004). So the Heideggerian *Sichöffnende und Offene* is perhaps not amenable to programming. To say it with Heideggerian overtones: programing can only be ontic but not ontological, since if the ontological were susceptible to programing, it would not be ontological. Agre has attempted to program routine activities and in doing so he has pragmaticized Dasein. But Heidegger himself warned specifically against this line of construing his philosophy, which reduces it to mere practical everyday activity:

*I attempted in Being and Time to provide a preliminary characterization of the phenomenon of world by interpreting the way in which we at first and for the most part move about in our everyday world. There I took my departure from what lies to hand in the everyday realm, from those things that we use and pursue, indeed in such a way that we do not really know of the peculiar character proper to such activity, and when we try to describe it we immediately misinterpret it by applying concepts and questions that have their source elsewhere. That which is so close and intelligible to us in our everyday dealings is actually and fundamentally remote and unintelligible to us. In and through this initial characterization of the phenomenon of world the task is to press on and point out the phenomenon of world as a problem. It never occurred to me, however, to try and claim or prove with this interpretation that the essence of man consists in the fact that he knows how to handle knives and forks or use the tram. (GA 29/30, pp. 262-263)*

In conclusion, it must be noted that the reception of Heidegger's philosophy in cognitive science, in particular, and the use of phenomenological notions for enriching the cognitive landscape, in general, is not merely the putting into work of those insights but more often than not also a *translation* of those terms into cognitive ones.

With this translation, something fundamentally phenomenological gets lost. Agre's work here discussed is a prominent example of how Heidegger's reception is rather 'analytic': 'analytic' not only in the sense that Heideggerian philosophy is appropriated by analytic-trained Anglo-American philosophers or, in this case, by an MIT-trained engineer, but also in the decisive sense that the Heideggerian philosophy which is appropriated for the purposes of advancing the new paradigm of cognition, pays only attention to specific parts of Division I of *Sein und Zeit*; parts which, in the same vein, are also appropriated very selectively. The reception is 'analytic' in that it constitutes a very schematic version of Heidegger taking precisely his thought out of context (Rehberg 2012, p. 160).

However, Agre's work is full of important insights, the most important of which is his demand that technical practice be aware of the philosophical metaphors pervading research in technological implementation.

[1] The Red Rock Eater News Service was a popular electronic mail service organized by Agre that was active in the mid 1990s, credited as a model for many of today's political blogs and online newsletters. See online: <http://polaris.gseis.ucla.edu/pagre/rre.html>. Retrieved August 20, 2014.

[2] In the informal essay, 'Critical Thinking for Technical People,' (originally an email message for the subscribers of the Red Rock Eater News Service), Agre tells the story "about how I became (relatively speaking, and in a small way) a better person through philosophy."

[3] It must be noted that, for the sake of the argument, reference is here made to early Blumenberg (2010) and his project of tracing absolute metaphors as those *Grundbestände* that cannot be conceptually reduced (*nicht in Begrifflichkeit aufgelöst werden können*) but rather function as constituting a catalytic sphere from which the universe of concepts continually renews itself.